# Super-resolution and non-diffraction longitudinal polarized beam

Haifeng Wang[1], L. P. Shi[1], B. S. Luk`yanchuk[1], C. J. R. Sheppard[2], T.C. Chong[1], G. Q. Yuan[1]

[1]*Data Storage Institute, DSI Building, 5 Engineering Drive 1, Singapore 117608*
[2]*Division of Bioengineering, National University of Singapore, Singapore 117576*

A method is presented for generation of a subwavelength (0.43λ) longitudinally polarized beam, which propagates without divergence over lengths of about 4λ in free space. This is achieved by controlling the amplitude, phase and polarization property of the Bessel-Gaussian field on the aperture of a high numerical aperture focusing lens.

A plane electromagnetic wave is purely transversal. Thus, it was for many years assumed impossible to create longitudinally polarized light in free space. However, later it was shown that any beam of finite diameter has a longitudinal field component, even in a free space [1, 2]. A strong longitudinal component appears at the focal region of a tightly focused laser beam [2-5]. It also arises with focusing of radially polarized light [6-12]. Besides academic interest, this longitudinal field has many attractive applications, e.g. in particle acceleration [2, 6, 7, 13], fluorescent imaging [14], second harmonic generation [15-17] and Raman spectroscopy [18]. It can permit the achievement of higher resolution in z-polarized confocal fluorescence microscopy [19] and scattering scanning near-field optical microscopy [20].

The longitudinal field can be suppressed or enhanced by amplitude, polarization and phase modulation of the incident beam [21]. For example, a longitudinal field can be completely suppressed in an azimuthally polarized beam [10, 22]. Several methods to enhance the longitudinal field component have been suggested [6, 11, 21, 23, 24], however all of them have insufficient optical efficiency (on the level of a few percents) and non-uniform axial field strength. Conversion efficiency is an critical characteristic for real applications, such as the use of radially and azimuthally polarized laser radiation for material processing [25], put into practice soon after the development of effective methods for radial and azimuthal beam polarization.

However, the conversion efficiency is not the only parameter which one has to improve. In an "ideal" situation we want to have a "pure" longitudinally polarized beam in free space with superresolution (also called subdiffraction) and uniform along optical axis (a so-called non-diffracting beam). Although it seems to be rather questionable that this problem can be solved (to the best of our knowledge from the published papers), we want to demonstrate that it is possible to combine all of these contradictory requirements using methods of phase modulation, which we applied to the radially-polarized Bessel-Gaussian beam [26, 27]. Namely, calculations show that it is possible to reach 17.5% efficiency of conversion to a beam that is almost "purely" longitudinally polarized along the optic axis. The FWHM of the longitudinal component intensity profile is $0.4\lambda$ (indicating superresolution), and the depth of focus is about $4\lambda$, i.e. we have a nondiffracting beam.

A schematic diagram of the suggested method is shown in Fig. 1. The Bessel-Gaussian beam undergoes a special phase modulation and is focused further by a high numerical aperture focusing lens. In the focal region we create a beam with the characteristics mentioned above.

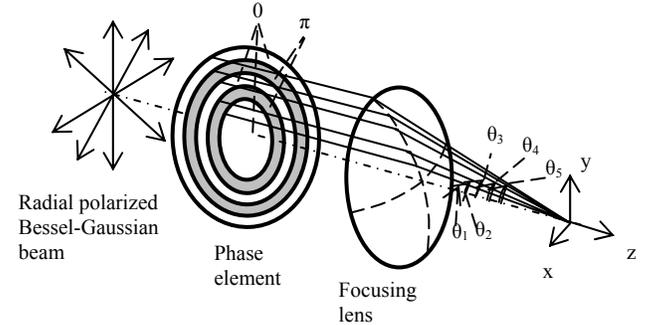

Fig. 1. Schematic diagram of the setup, phase modulation optical element and focusing lens.

First we review some effects of amplitude and phase modulation related to superresolution and nondiffracting beam problems. The amplitude modulation of the incident beam follows from the analysis of non-diffracting beams. This idea comes from the exact solution of Maxwell equations in terms of cylindrical waves. The zero order Bessel beam mode, with amplitude $\propto J_0(k_r r)\exp(-ik_z z)$ propagates in a free space without diffraction [3, 6-8, 23, 28]. This kind of beam exists in infinite free space and has polarization perpendicular to the propagation direction in the paraxial approximation. Although the zero Bessel function is not square integrable and, thus, an infinite power is necessary to create a nondiffracting beam, some "nearly nondiffracting" beams (e.g. Bessel-Gaussian, Laguerre-Gaussian, Hermite-Gaussian etc., see [26, 29-31]) with finite power can be realized and they can propagate over a long range without significant divergence.

Following the general theory of polarized beam focusing [1] one can write for the electric fields near focus $z=0$ for illumination of the high aperture lens with the waist of a radially polarized Bessel-Gaussian beam, see e.g. [6, 8, 27]

$$E_r(r,z) = A\int_0^\alpha \cos^{1/2}\theta \sin(2\theta)\ell(\theta)J_1(kr\sin\theta)e^{ikz\cos\theta}d\theta, \quad (1)$$

$$E_z(r,z) = 2iA\int_0^\alpha \cos^{1/2}\theta \sin^2\theta \,\ell(\theta)J_0(kr\sin\theta)e^{ikz\cos\theta}d\theta. \quad (2)$$

Here we have adopted notations similar to [27]: $\alpha = \arcsin(NA/n)$, where NA is the numerical aperture and $n$ is the index of refraction between the lens and the sample, $J_0(x)$ and $J_1(x)$ denote Bessel functions and the function $\ell(\theta)$ describes amplitude modulation. For illumination by a Bessel-Gaussian beam this function is given by [27]:

$$\ell(\theta) = \exp\left[-\beta^2\left(\frac{\sin\theta}{\sin\alpha}\right)^2\right]J_1\left(2\gamma\frac{\sin\theta}{\sin\alpha}\right), \quad (3)$$

where $\beta$ and $\gamma$ are parameters that in our calculations we take as unity. Illumination with a Laguerre-Gauss beam is very similar. We also assume $n=1$ and a high numerical aperture $NA = 0.95$ (corresponding to $\alpha \approx 71.8°$). The corresponding field distribution is shown in Fig. 2.

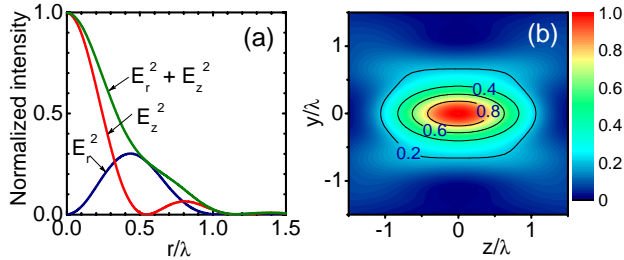

Fig. 2 (color online). a) Intensity profile of the radial component, longitudinal component and the total field on the focal plane of the NA = 0.95 lens for radial polarized Bessel-Gaussian beam. b) Contour plot for the total intensity distribution.

One can see from Fig. 2 that the intensity of the longitudinal field is high, but the "parasite" radial field intensity consists of about 30% of the total intensity. This radial field leads also to a broadening of the total intensity distribution. As a result the total intensity spot size is as big as FWHM = 0.68 $\lambda$, which is larger than the diffraction limit of this focusing lens $\lambda/2(NA) = 0.526\ \lambda$. The intensity spot size of the longitudinal field component is only 0.49 $\lambda$. This "parasite" influence of the radial field intensity is quite a general effect which one can see with other generalized Bessel-Gaussian beams also, e.g. in [29]. Thus, to have a "good" longitudinally polarized beam one should suppress the radial field component. We show that this is possible to do with additional phase modulation.

It is well-known that the beam can be modulated using a phase-mask element. Formally it corresponds to replacing the function $\ell(\theta)$ by the function $T(\theta) = \exp[i\varphi(\theta)]$ in Eqns. (1), (2). If one sets the element with $\varphi(\theta) = 0$ or $\varphi(\theta) = \pi$ for different ranges of angle $\theta$ it corresponds to a so-called binary mask [32]. Phase modulation by itself (without additional amplitude modulation) permits the creation of "nearly nondiffracting" beams. For example the seven belt $\pi$-phase binary optical element placed on the aperture of a high numerical aperture (NA = 0.85) aplanatic focusing lens produces a subwavelength super-resolution 0.42 $\lambda$ light beam which propagates over a long distance ($z \approx 5\lambda$) without divergence [33]. However, light within the focal range is substantially transversally polarized, the field along the y-direction of focal spot does not contain longitudinal component at all, and no super-resolution effect can be observed in the x-direction, because the parasite longitudinal field makes the total intensity spot larger.

Now we want to apply additional phase modulation to the radial polarized Bessel-Gaussian beam to suppress the radial component in the focal region. Formally it corresponds to replacing the function $\ell(\theta)$ by the function $\ell(\theta)T(\theta)$ where we assume a five belt optical element with

$$T(\theta) = \begin{cases} 1, & \text{for } 0 \leq \theta < \theta_1, \theta_2 \leq \theta < \theta_3, \theta_4 \leq \theta < \alpha, \\ -1, & \text{for } \theta_1 \leq \theta < \theta_2, \theta_3 \leq \theta < \theta_4. \end{cases} \quad (4)$$

The four angles $\theta_i\ (i=1,...,4)$ were optimized to maximize the ratio of the longitudinal to the radial component in the focal cross-section $z=0$. The set of angles obtained are

$$\theta_1 = 4.96°, \ \theta_2 = 20.68°, \ \theta_3 = 33.03°, \ \theta_4 = 45.88°. \quad (5)$$

The general modulation function $\ell(\theta)T(\theta)$ on the aperture of the focusing lens for this case is shown in Fig. 3(a). The intensity profiles of the radial component, the longitudinal component and total field of the longitudinally polarized beam in the focal cross-section are shown in Fig. 3(b). The FWHM of the longitudinal component intensity profile in Fig. 3(b) is 0.4 $\lambda$, which is smaller than that shown in Fig. 2(a) (0.49 $\lambda$). The FWHM of the total intensity spot in Fig. 3(b) is 0.43 $\lambda$, so the spot area is 0.15 $\lambda^2$, which is smaller than that obtained without the binary optical element shown in Fig. 2(a) (FWHM = 0.68 $\lambda$), where the spot area is 0.36 $\lambda^2$. The *super-resolution* effect results in a spot area reduced by 58%. The efficiency of conversion from a radially-polarized

Bessel-Gaussian beam to the beam with longitudinal polarization is about 17.5%. A contour plot of the total intensity image is shown in Fig. 3(c). One can see that the total depth of focus is about $4\lambda$ with uniform axial intensity, i.e. we have close to a nondiffracting beam. The modulation function produces a uniform axial intensity over a finite range, which is different from that generated by a sinc variation [34].

The radially polarized field (1), (2) produces an azimuthally polarized magnetic field with an $H_\varphi$ component only (similar to the field around a linear conductor carrying current). This field can be found from Maxwell's equations, yielding $H_\varphi = -i(\partial_z E_r - \partial_r E_z)/k$, i.e.

$$H_\varphi = 2A\int_0^\alpha \cos^{1/2}\theta \sin(\theta)\ell(\theta)T(\theta)J_1(kr\sin\theta)e^{ikz\cos\theta}d\theta. \quad (6)$$

The distribution of magnetic field component $H_\varphi$ is shown in Fig. 3(d). In the focal plane $z = 0$ the field (6) is purely real. Different signs of this field correspond to clockwise and counter-clockwise field directions. In Fig. 3(d) we present both the Bessel-Gaussian field and the corresponding field after phase modulation in the focal plane.

In principle it is possible to do further enhancement of the longitudinal field by using a higher numerical aperture lens. However with NA > 1, the field diverges subsequently away from the focal plane, thus the nondiffraction characteristic of the beam is lost. Also it is possible to improve the longitudinal beam quality (and obtain longer depth of focus) by applying more belts, e.g. seven belts instead of the five belts in Eq. (4). However efficiency of optical conversion also drops with increasing the total number of belts. Thus, the five-belt optical modulation is a compromise between the beam quality and optical efficiency. For this case, the conversion efficiency is about 17.5% and the beam quality (ratio of the longitudinally polarized field intensity to the total intensity within the directional lobe of the focused beam) is above 95 %.

Once again we want to emphasize that we are talking about formation of uniform, longitudinally polarized light along an axis in free space, where the vanishing of one field component (necessary from the uncertainty principle to overcome the diffraction limit) is created by the high aperture lens. It is different from the near field region between two different materials with evanescent or plasmonic waves. The necessary conditions for the field localization in the latter case are fulfilled in a "natural way" due to exponential intensity attenuation. It has been demonstrated that it is possible to create radially polarized [35] or longitudinally polarized light [36] near the surface of metal with the help of resonant excitation of surface plasmon waves.

Some comments should be given concerning the Poynting vector field of the longitudinally polarized beam, as shown in Fig. 4. The time-averaged Poynting vector is determined by

$$\langle \mathbf{S} \rangle = \frac{c}{4\pi}\text{Re}(\mathbf{E}\times\mathbf{H}^*), \quad (7)$$

where the asterisk denotes the operation of complex conjugation. This field is axially symmetric because of the condition $S_\varphi = 0$. Thus, the field lines of the Poynting vector follow the equation $dr/dz = S_r/S_z$. Except on the singular line $r = 0$, the Poynting vector field contains circular singular lines, which are distributed in the focal plane $z = 0$. Cross-section of these lines by the $yz$-plane corresponds to the singular points 1 and 2 in Fig. 4. In principle, such types of singularities of the Poynting vector for focused beam are well-known [3, 37]. Similar vortices arise also during scattering of light on nanoparticles and nanowires near plasmon resonance frequencies [35, 38, 39]. However, in contrast to these papers, here these vortices arise with toroidal symmetry. Toroidal symmetry of the Poynting field vortices results in the formation of the longitudinal field polarization. The mechanism of this effect is rather simple. Because magnetic field consists of an $H_\varphi$ component only, it means that the electric field is directed perpendicularly to the Poynting vector lines. As a result this toroidal Poynting vector "hoop" creates a radial field, which compensates the radial component of the electric field inside the hoop. Thus, the polarization of the field near to the axis becomes "more longitudinal".

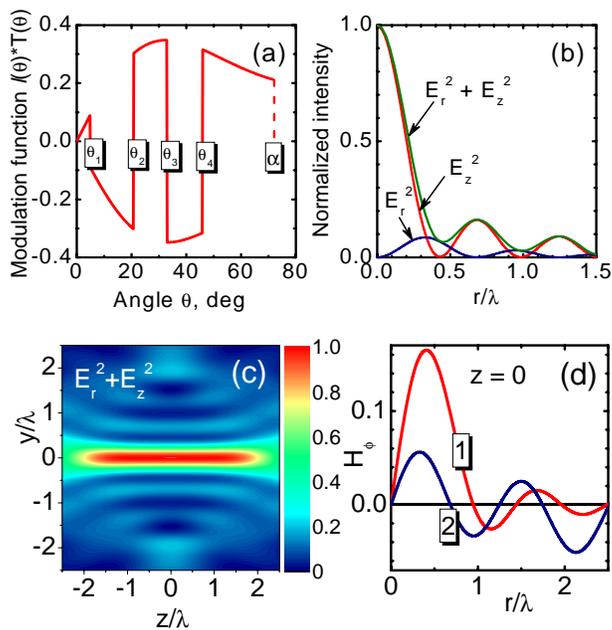

Fig. 3 (color online). a) Modulation function $\ell(\theta)T(\theta)$ according to Eqns. (3)-(5). b) Intensity profile of the radial component, longitudinal component and the total field on the focal plane of the NA = 0.95 lens for radial polarized Bessel-Gaussian beam with additional phase modulation. c) Contour plot for the total intensity distribution in $yz$ plane. d) Radial distribution of magnetic field $H_\varphi$ at focal plane $z = 0$. Curve 1 presents Bessel-Gaussian field and curve 2 the magnetic field after phase modulation.

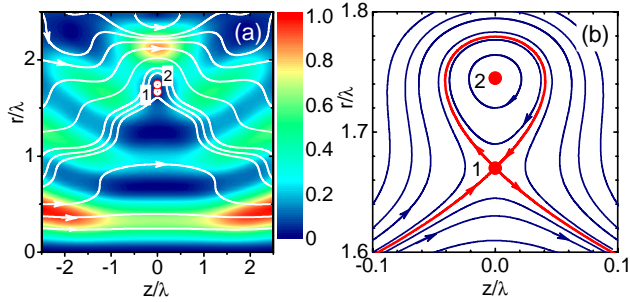

Fig. 4 (color online). a) Color density plots show values of modulus of the Poynting vector normalized over its maximal value in the figure. Field lines are shown in white and singular circular lines as open red circle. b) The "zoom" for the energy flow in the vicinity of singular points 1 and 2. The thick red line indicates separatrix, which crossed the saddle point 1. Another singular point 2 is of the "center" type. Vortices arise around of this "center" inside the loop of separatrix. In 3D geometry they present toroidal vortices.

It also works in a similar way to the focusing of electron beam with magnetic lens, which may suggest a possible way to control photons.

In conclusion, we have shown that, longitudinal field, which is usually taken as static, can exist in the form of propagating beam like linear or circular polarized beam, and this beam can propagate without divergence for over 4 $\lambda$ (nondiffracting beam). It is also highly localized in transversal direction with FWHM = 0.43 $\lambda$ (superresolution). It looks like a needle as is shown in Fig. 3c. However when a beam size is smaller than half wavelength, it usually diverge in all directions with intensity attenuate exponentially in free space [40].

--------------------

4